\documentstyle[twocolumn,prb,aps]{revtex}
\begin{document}
\draft
\title{Dynamic Magneto-Conductance Fluctuations and\\
Oscillations in Mesoscopic Wires and Rings}
\author{D.\ Z.\ Liu,$^1$ Ben Yu-Kuang Hu,$^{1,2}$ C.\ A.\ Stafford,$^1$ and
S.\ Das Sarma$^1$}
\address{$^1$ Department of Physics,
University of Maryland,
College Park, Maryland 20742-4111}
\address{$^2$ Mikroelektronik Centret, Bygning 345 \O, Danmarks Tekniske
Universitet, DK-2800 Lyngby, Denmark}
\date{\today}
\maketitle

\makeatletter
\global\@specialpagefalse
\def\@oddhead{REV\TeX{} 3.0\hfill Das Sarma Group Preprint, 1994}
\let\@evenhead\@oddhead
\makeatother

\begin{abstract}
Using a finite-frequency recursive Green's function technique,
we calculate the dynamic magneto-conductance fluctuations and
oscillations in disordered mesoscopic normal metal systems,
incorporating inter-particle Coulomb interactions
within a self-consistent potential method.
In a disordered metal wire, we observe ergodic behavior
in the dynamic conductance fluctuations.
At low $\omega$, the real
part of the conductance fluctuations is essentially given by the dc
universal conductance fluctuations while the imaginary part increases
linearly from zero, but for $\omega$ greater than the Thouless energy
and temperature, the fluctuations decrease as $\omega^{-1/2}$.
Similar frequency-dependent behavior is found for
the Aharonov-Bohm oscillations in a metal ring.
However, the Al'tshuler-Aronov-Spivak oscillations, which predominate at
high temperatures or in rings with many channels, are strongly
suppressed at high frequencies, leading to interesting crossover
effects in the $\omega$-dependence of the magneto-conductance oscillations.

\end{abstract}

Quantum transport in mesoscopic normal metal
systems has been extensively studied both experimentally and
theoretically in the last
decade.\cite{lee,stone1,aas,aak,stone2,buttiker1,stone3,xie,buttiker2,webb}
Most of the work has focused on
interference effects in {\em static} ($\omega=0$) transport properties,
such as universal conductance
fluctuations (UCF) of order $e^2/h$ in wires and $hc/e$-periodic Aharonov-Bohm
oscillations in mesoscopic rings.  Recently, dynamic
properties of these systems have begun to be
investigated.\cite{fu,buttiker3,jauho,pieper}
The dynamic response
of mesoscopic systems is of fundamental interest because the frequency
introduces another energy scale into the problem which plays a role
quite different from temperature, and there are possible
device applications, such as a mesoscopic photovoltaic
effect device.\cite{falko}

In this {\it rapid communication}, we present a theoretical investigation
of dynamic magneto-conductance fluctuations and oscillations in disordered
mesoscopic normal metal systems, using the recently developed
finite-frequency Landauer-B\"{u}ttiker formalism.\cite{fu,buttiker3}
An advantage of this formalism is that the qualitative frequency dependence
can be understood by simple arguments.
Furthermore, this method can take into account the internal potential
distribution in the sample due to the dynamic response of the system by
an approximate self-consistent potential method.\cite{buttiker3}

We consider the diffusive but phase-coherent transport regime, where the
elastic mean-free path is less than the sample length $L$, but
the temperature
$T$ is low enough so that the inelastic mean-free path is much larger
than $L$.
There are then three important energy scales
in the problem:  $k_BT$, $\hbar\omega$, and the Thouless energy $E_c$.
$E_c$ is defined\cite{stone2} as the energy change necessary for
electrons traversing the sample to pick up a phase difference of order unity.
For diffusive transport, where the typical path length $S \sim v_F L^2/D$,
\begin{equation}
E_c\simeq hv_F/S \simeq hD/L^2 \simeq g_c\Delta E,
\end{equation}
where $D$ is the diffusion constant, $\Delta E$ is the level
spacing at $E_F$, and
$g_c$ is the dc conductance (in units of $e^2/h$).
Interesting effects occur due to the interplay
of these three energy scales.

We find the following results:
(1) In disordered wires at $T=0$, both the real and (for sufficiently large
frequencies) imaginary parts of the conductance fluctuations
$\delta g(\omega)$ show UCF for $\omega < E_c$, and both
fall as $\omega^{-1/2}$ for $\omega \gg E_c$.  These fluctuations are
ergodic over all frequencies investigated.  (2) In disordered metal
rings, the $hc/e$ Aharonov-Bohm (AB)
conductance oscillations
with respect to magnetic field persist to high frequencies, and
have a frequency dependence similar to that of the conductance
fluctuations.
The $hc/2e$ Altshuler-Aronov-Spivak (AAS) oscillations, which
predominate at high temperatures or in samples with many channels,
show a more rapid decrease with frequency;\cite{aas} as a result,
the magneto-conductance oscillations may cross over from AAS to AB behavior
when the frequency is increased.

Following conventional treatments,\cite{landauer} we consider the
mesoscopic sample as a disordered region which scatters electrons
incident from semi-infinite ordered regions (perfect ``leads'').
Using the recently developed finite-frequency Landauer-B\"{u}ttiker formula,
the conductance matrix for a multi-probe system of noninteracting
electrons, $g_{\alpha\beta}(\omega) \equiv
<\delta I_{\alpha}(\omega)>/\delta U_{\beta}(\omega)$,
where $<I_\alpha>$ is the current in lead $\alpha$
and $U_{\beta}$ is the potential in lead $\beta$,
may be written as\cite{fu,buttiker3}
\begin{eqnarray}
 g_{\alpha\beta}(\omega)
    & = & \frac{e^2}{h}\int dE\,\mbox{Tr}
    \{{\bf 1}_{\alpha}\delta_{\alpha\beta}
     -{\bf s}_{\alpha\beta}^{\dagger}(E){\bf s}_{\alpha\beta}(E+\hbar\omega)\}
\label{eq3}
\\
& &\mbox{ }\mbox{ }\mbox{ }\mbox{ } \times
\frac{f(E)-f(E+\hbar\omega)}{\hbar\omega}, \nonumber
\end{eqnarray}
where ${\bf s}_{\alpha\beta}(E)$ is the scattering matrix for electrons of
energy $E$, and the trace over channels includes spin.
Eq.\ (\ref{eq3}) is valid\cite{fu,buttiker3} for $\omega \ll 2 \pi v_F/L$.
Thus the dynamic conductance is related to
the correlation of transmission (or reflection) coefficients
at energies differing by $\hbar\omega$.
For a two-lead non-interacting mesoscopic system ({\it i.e.} $\alpha,
\beta=1,2$) with a symmetric drop in potential across the device,
the measured conductance is $g(\omega)=[g_{11}(\omega)-g_{12}(\omega)]/2$.

We simulate the mesoscopic system using a tight-binding
model on a square lattice, with unit nearest-neighbor hopping matrix
element.  The disorder is modeled by an on-site
random potential ranging from $-W/2$ to $W/2$.
A magnetic field modifies the hopping term by a Peierls phase factor.
The retarded
single particle Green's function $G^+$ is calculated using the
recursive Green's function algorithm.\cite{lee2,mackinnon}
For an $L\times M$ system (length $L$, width $M$),
the transmission and reflection
coefficients ${\bf t} \equiv {\bf s}_{12}$ and ${\bf r} \equiv {\bf
s}_{11}$ are determined using the following relations:\cite{stone1}
\begin{mathletters}
\begin{eqnarray}
r_{mn}&=&i(v_mv_n)^{1/2}G_{mn}^{+}(0,0)-\delta_{mn},  \\
t_{mn}&=&i(v_mv_n)^{1/2}G_{mn}^{+}(0,L)e^{ik_nL},
\end{eqnarray}
\end{mathletters}
\noindent
where $m,n=1,2,...,M$ is the channel number, $v_m$ is the group
velocity for channel $m$, and $k_n$ is the wavevector for channel $n$.
In Fig.~1, we show the real (solid curve) and imaginary (dashed curve)
parts of the dynamic
conductance of a $100 \times 4$ wire versus frequency $\omega$,
calculated for noninteracting electrons using Eqs.~(\ref{eq3}) and (3).
Fig.~1 illustrates the generic frequency dependence of a diffusive
metallic conductor:  $|\mbox{Im}[g]|$ increases
linearly for small frequencies, becoming comparable to $e^2/h$ when
$\omega \sim E_c$; $\mbox{Re}[g]$ exhibits a weak-localization
suppression at low frequencies; both exhibit random fluctuations of
order $e^2/h$ on a frequency scale $\mbox{} \sim E_c$, which
decrease in amplitude with increasing frequency.
It should be mentioned that the results shown in Fig.\ 1 at frequencies
greater than $\omega_{\rm max} = 2 \pi v_F/L = 0.13$ are, strictly
speaking, beyond the range of validity of Eq.\ (\ref{eq3}).  However,
since ${\bf s}_{\alpha \beta}(E)$ has no structure at this frequency
scale for a diffusive metallic conductor, we believe these results are
representative for the somewhat larger systems studied
experimentally,\cite{webb,pieper} for which the Thouless energy and the
inverse ballistic transit time are separated by several orders of
magnitude.

The dynamic response of an {\em interacting}
mesoscopic system depends in detail on the inhomogeneities in the local
electric field, which must be determined self-consistently
from the external potentials and the dynamic charge distribution in the
sample, in contrast to the dc case, where inhomogeneities in the local
electric field are irrelevant in linear response.
To lowest order, the self-consistent potential
due to the interactions in the system can be
approximated by a constant induced potential
resulting from the charge pileup in the system,
$U_0^{ind}=Q_0/C$, where $C$ is the effective capacitance of the
mesoscopic wire or ring.
Within this approximation, the
response of the interacting system $g_{\alpha\beta}^I$ can be determined
from the admittances $g_{\alpha\beta}$ for the non-interacting
system as\cite{buttiker3}
\begin{equation}
g_{\alpha\beta}^I(\omega)=g_{\alpha\beta}(\omega)-\frac{(i/\omega
C)\sum_{\gamma}g_{\alpha\gamma}(\omega)\sum_{\delta}
g_{\delta\beta}(\omega)}{1+(i/\omega C)
\sum_{\gamma\delta}g_{\gamma\delta}(\omega)}. \label{gi}
\end{equation}
In Fig.~1,
the dynamic conductance of the interacting system ($C=0$) is compared to
that of the noninteracting system ($C=\infty$).
As is evident from Fig.~1,
the corrections to the dynamic conductance due to the self-consistent
potential are proportional to $\omega$ for small
frequencies,\cite{buttiker3}
and are largest in
magnitude when $\omega \sim E_c$, where the fluctuations away from zero
of the sums over admittances in the second term of Eq.~(\ref{gi}) are
maximal.  The relevant capacitance scale determining the crossover from
strongly to weakly interacting behavior is set by the dwell time $\tau$
of an electron in the system:\cite{buttiker3}  if $R C \gg \tau$,
interactions can be neglected; if $R C \ll \tau$, interactions enforce
charge neutrality.
For a typical diffusive mesoscopic conductor, such as the Ag ring
of Ref.~\onlinecite{pieper},
the $RC$ time is estimated to be several orders of magnitude smaller
than the dwell time $\sim h/E_c$, so effectively $C=0$.
As shown in Fig.\ 1, the correction to the conductance due to
self-consistent charging effects can be considerable
(of order $e^2/h$).  However, we find {\em no} significant
change in the amplitude of the
conductance {\em fluctuations} or {\em oscillations}
when the capacitive effect is incorporated.
In the following, averaged quantities shown are for $C=\infty$, while
results for a single sample use $C=0$.

Conductance fluctuations similar to those shown as a function of
$\omega$ in Fig.\ 1 are obtained when the other parameters of the system
are varied.  In Fig.\ 2, we show
the dynamic magneto-conductance fluctuations
in disordered mesoscopic wires at $T=0$
calculated by averaging over (a) an
ensemble of samples, (b) magnetic field, and (c) chemical potential.
Regardless of the averaging method, the results obtained were equivalent
within our statistical uncertainty
(see Fig.\ 2), demonstrating the {\em ergodicity} of the system.
At low frequencies, $\omega\ll E_c$,
the root mean square value of the real part of the conductance fluctuations,
$\mbox{Re}[\Delta g]$, is given by the UCF value ($\approx 0.6e^2/h$),
independent of sample size (as long as the sample is in the quantum
coherent transport regime, as we have assumed),
while $\mbox{Im}[\Delta g]$ increases linearly
and saturates at the same universal value.
In the high frequency limit, $\omega \gg E_c$,
$\Delta g \sim \omega^{-1/2}$, as shown in Fig.\ 2.
The $\omega$-dependence of $\Delta g$ may be understood by simple
qualitative arguments based on Eq.~(\ref{eq3}):
$g_{\alpha\beta}$ is real for $\omega = 0$, but the product
$s_{\alpha\beta}^{\dagger}(E) s_{\alpha\beta}(E+\hbar\omega)$
(for a single channel) acquires a complex phase of order unity when
$\hbar\omega \sim E_c$, so $\mbox{Re}[\Delta g]$ and $\mbox{Im}[\Delta
g]$ become comparable at that frequency.  For $\hbar \omega \gg E_c$,
the product $s_{\alpha\beta}^{\dagger}(E) s_{\alpha\beta}(E+\hbar\omega)$
has an arbitrary complex phase which varies by an amount of order unity
when $E \rightarrow E+E_c$.  Standard random walk arguments applied to
the integral in Eq.~(\ref{eq3}), whose range is roughly from $\mu -
\omega$ to $\mu$, then give $\mbox{Re}[\Delta g],\, \mbox{Im}[\Delta g]
\sim (E_c/\omega)^{1/2}$ for $\hbar \omega \gg E_c,\, k_B T$.

We now consider dynamic magneto-transport in mesoscopic rings.
In the dc case, it is well known that a magnetic flux placed
through a mesoscopic ring generates periodic conductance oscillations
with flux period $hc/e$ [the Aharonov-Bohm (AB) effect], and $hc/2e$
[the Al'tshuler-Aronov-Spivak (AAS) effect].
Both have been observed experimentally in mesoscopic
rings.\cite{aas,stone2,webb}
In Fig.\ 3, we show the calculated conductance oscillations
in a mesoscopic ring of circumference 80 with 4 transverse channels
(denoted $40 \times 4$)
at $\omega \sim E_c$ and $T=0$.  For these parameters, the real and
imaginary parts of the AB amplitude are comparable, and much greater
than the AAS amplitude.
The root mean square AB and AAS amplitudes
for an ensemble of 20 $100 \times 4$ rings at $T=0$ are shown in Fig.\
4(a).  The frequency dependence of the AB amplitude\cite{pieper2}
is similar to that
of the conductance fluctuations (Fig.\ 2),
while the AAS amplitude is smaller, and
has a more rapid decrease with frequency.  Qualitatively, this
$\omega$-dependence may be understood from Eq.~(\ref{eq3}):  The AB
effect comes from the quantum interference of electrons going through
opposite branches of the ring; since these paths have a random phase
difference in zero field even at $\omega = 0$, it is not important that
${\bf s}_{\alpha\beta}$ and ${\bf s}_{\alpha\beta}^{\dagger}$
are evaluated at different energies in Eq.~(2) (except that this makes
the AB amplitude complex), and the frequency
dependence of the AB effect is therefore similar to that of the conductance
fluctuations.  However, the AAS effect comes from the interference of
time-reversed paths encircling the flux which have equal phases
in zero field, so that the AAS effect is insensitive to energy
averaging at $\omega = 0$.  But the finite frequency electric field
breaks the time-reversal symmetry of the paths contributing to the AAS
effect,\cite{aak} leading to an $\omega^{-1/2} \exp[-(\omega/E_c)^{1/2}]$
suppression of AAS oscillations at high frequencies.\cite{aas}
The different frequency
dependence of the AB and AAS amplitudes is particularly evident in Fig.\
4(b), which is for a single sample at $T=0.05 > E_c$.  The finite
temperature suppresses the AB effect by energy averaging, but not the
AAS effect, so that the two are comparable at $\omega = 0$.  As the
frequency is increased, the AAS effect is strongly suppressed, but the
AB effect is not expected to be suppressed until $\omega > k_B T$, and
appears to be constant (up to fluctuations) in Fig.\ 4(b).
This crossover from AAS oscillations to AB oscillations demonstrates
the qualitatively  different role of temperature and frequency in
quantum transport in mesoscopic systems.\cite{mondo}
We point out that the $\omega$-dependence of the AB effect shown in
Fig.\ 4(b) is consistent with the
experimental results in Ref. \onlinecite{pieper}.  However, we find that
the smallness of $\mbox{Im}[A]$ relative to $\mbox{Re}[A]$ at
frequencies greater than $E_c$ is sample specific, and does not persist
in the ensemble average [Fig.\ 4(a)].

In summary, we have calculated
the dynamic magneto-conductance for disordered mesoscopic normal metal
wires and rings.
In a metal wire, we find that the dynamic conductance fluctuations
decrease as $(E_c/\omega)^{1/2}$ when the frequency is much larger than the
Thouless energy and temperature; at lower frequency, the real
part of the conductance fluctuations is essentially given by the dc
universal conductance fluctuations while the imaginary part increases
linearly from zero. Similar frequency-dependent behavior is found for
the $hc/e$-periodic Aharonov-Bohm
oscillation amplitude in a metal ring.
For high enough temperatures, we find
that the oscillations may cross over
from $hc/2e$ period to $hc/e$ period with increasing $\omega$.
We find that incorporating interactions through a self-consistent
potential changes the magneto-conductance of an individual
mesocopic conductor considerably, but does not affect the magnitude of
the conductance fluctuations and oscillations.

We acknowledge M.\ B\"{u}ttiker for stimulating discussions and
valuable advice.  We also thank T.\ Kawamura, J.\ Pieper and
J.\ Price for helpful
discussions.  This work was supported by the National Science Foundation
and the U.S.\ Office of Naval Research.



\begin{figure}
 \vbox to 5.0cm {\vss\hbox to 6cm
 {\hss\
   {\includegraphics{fig/f1.ps}
   }
  \hss}
 }
\caption{Dynamic conductance of a $100\times 4$ wire
versus frequency $\omega$.  Both
real (solid line for $C=\infty $, dash dot line for $C=0$) and
imaginary (dashed line for $C=\infty $, dotted line for $C=0$)
parts of the conductance are shown.  The disorder amplitude is $W=1$,
and $E_c$ is estimated to be $0.03$.
}
\end{figure}

\begin{figure}
 \vbox to 5.0cm {\vss\hbox to 6cm
 {\hss\
   {\includegraphics{fig/f3.ps}
   }
  \hss}
 }
\caption{Frequency-dependence of the conductance fluctuations for
$100\times 8$ wires. Both real (filled
symbols connected by solid line) and imaginary (open symbols connected
by dotted line) are shown. Cirles are for 100 sample
ensemble average; Squares are for magnetic
field average ($\alpha=0\sim 0.1$); Triangles are for
chemical potential avegrage ($\mu=-0.4\sim 0.4$).
The dashed line indicates the $\omega^{-1/2}$ behavior.
$W=1$ and $E_c \simeq 0.03$.
}
\end{figure}

\begin{figure}
 \vbox to 5.0cm {\vss\hbox to 6cm
 {\hss\
   {\includegraphics{fig/f4.ps}
   }
  \hss}
 }
\caption{The Aharonov-Bohm effect for a $40\times 4$ ring at finite
frequency ($\omega=0.02 \sim E_c$). Here the real part (solid line) is offset
by
$-4e^2/h$ to compare with the imaginary part (dotted line).
The field in the anulus is 25\% of that in the hole.
}
\end{figure}

\begin{figure}
 \vbox to 5.0cm {\vss\hbox to 6cm
 {\hss\
   {\includegraphics{fig/f5a.ps}
   }
  \hss}
 }
 \vbox to 5.0cm {\vss\hbox to 6cm
 {\hss\
   {\includegraphics{fig/f5b.ps}
   }
  \hss}
 }
\label{f5}
\caption{Frequency dependence of magneto-conductance oscillation
amplitudes in mesoscopic rings. $W=1$ and $E_c \simeq 0.01$.
The circle is for the AB oscillations,
triangle for the AAS oscillations; The solid symbol is for the real
part, open symbol for the imaginary part. (a) Zero temperature, 20 sample
ensemble average.  The
dashed line indicates $\omega^{-1/2}$ behavior;
 (b) $T=0.05$, single sample, $C=0$.
}
\end{figure}


\begin{references}
\bibitem{lee} P. A. Lee, Physica {\bf 140A}, 169 (1986); P. A. Lee and
A. D. Stone, Phys. Rev. Lett. {\bf 55}, 1622 (1985).
\bibitem{stone1} A. D. Stone, Phys. Rev. Lett. {\bf 54}, 2692 (1985)
and references therein.
\bibitem{aas} B. L. Al'tshuler, A. G. Aronov, and B. Z. Spivak, JETP
lett. {\bf 33}, 94 (1981).
\bibitem{aak} B. L. Al'tshuler, A. G. Aronov, and D. E. Khmelnitsky,
Solid State Commun. {\bf 39}, 619 (1981).
\bibitem{stone2} A. D. Stone and Y. Imry, Phys. Rev. Lett. {\bf 56},
189 (1986) and references therein.
\bibitem{buttiker1} M. B\"{u}ttiker, Phys. Rev. Lett. {\bf 57}, 1761 (1986).
\bibitem{stone3} H. U. Baranger, A. D. Stone, and D. P. DiVincenzo, Phys.
Rev. B {\bf 37}, 6521 (1988).
\bibitem{xie} X. C. Xie and S. Das Sarma, Phys.
Rev. B {\bf 38}, 3529 (1988).
\bibitem{buttiker2} M. B\"{u}ttiker, Phys. Rev. Lett. {\bf 65}, 2901 (1990).
\bibitem{webb} R. A. Webb, S. Washburn, C. P. Umbach, and R. B.
Laibowitz,  Phys. Rev. Lett. {\bf 54}, 2696 (1985).
\bibitem{fu} Y. Fu and S. C. Dudley, Phys. Rev. Lett. {\bf 70}, 65 (1993).
\bibitem{buttiker3} M. B\"{u}ttiker, A. Pr\^{e}tre, and H. Thomas, Phys.
Rev. Lett. {\bf 70}, 4114 (1993); {\bf 71}, 465 (1993).
\bibitem{jauho} Ned S. Wingreen, Antti-Pekka Jauho and Yigal Meir,
Phys.\ Rev.\ B {\bf 48}, 8487 (1993).
\bibitem{pieper} J. B. Pieper and J. C. Price, Physica B {\bf 194/196},
1051 (1994).
\bibitem{falko} V. Fal'ko, Europhys. Lett. {\bf 8}, 785 (1989).
\bibitem{landauer} R. Landauer, Philos. Mag. {\bf 21}, 863 (1970);
D. Fisher and P. A. Lee, Phys. Rev. B {\bf 23}, 6851 (1981);
P. A. Lee and D. Fisher, Phys. Rev. Lett. {\bf 47}, 882 (1981).
\bibitem{lee2} S. Das Sarma and Song He, Int. J. Mod. Phys. B {\bf
7}, 3375 (1993) and references therein.
\bibitem{mackinnon} A. Mackinnon, J. Phys. C {\bf 13}, L1031 (1980);
A. Mackinnon, Z. Phys. B {\bf 59}, 385 (1985).
\bibitem{pieper2} A similar frequency dependence for the AB effect
was obtained by a different method by
J. B. Pieper and J. C. Price, preprint (unpublished).  However, they did not
consider charging effects.
\bibitem{mondo}
A similar crossover is expected at $T=0$
in a ring with many transverse channels, in which
the AAS effect is expected to be dominant at low frequencies.
\end{references}
\end{document}